\documentclass[doublecol,figures, a4paper, final]{epl2}

\title{The ground states of the two-component order parameter superconductor}
\shorttitle{Ground state for the two-component superconductor} 

\author{Mauro M. Doria\inst{1} \and Antonio R. de C. Romaguera\inst{2} \and F. M. Peeters\inst{3,4}}
\shortauthor{Mauro M. Doria \etal}

\institute{
  \inst{1} Departamento de F\'{\i}sica dos S\'{o}lidos, Universidade Federal do Rio de Janeiro, 21941-972 Rio de Janeiro, Brazil\\
  \inst{2} Departamento de F\'{\i}sica, Universidade Federal Rural de Pernambuco, 52171-900 Recife, Pernambuco, Brazil \\
  \inst{3}{Departamento de F\'{\i}sica, Universidade Federal do Cear\'a, 60455-760 Fortaleza, Cear\'a,
  Brazil}\\
  \inst{4} Departement Fysica, Universiteit Antwerpen, Groenenborgerlaan 171, B-2020 Antwerpen, Belgium
}

\abstract{ We show that in presence of an applied external field the
two-component order parameter superconductor falls in two categories
of ground states, namely, in the traditional Abrikosov ground state
or in a new ground state fitted to describe a superconducting layer
with texture, that is, patched regions separated by a
phase difference of $\pi$. The existence of these two kinds of
ground states follows from the sole assumption that the total
supercurrent is the sum of the two individual supercurrents and is
independent of any consideration about the free energy expansion.
Uniquely defined relations between the current density and the
superfluid density hold for these two ground states, which also
determine the magnetization in terms of average values of the order
parameters. Because these ground state conditions are also Bogomolny
equations we construct the free energy for the two-component
superconductor which admits the Bogomolny solution at a special
coupling value.}

\begin{document}

\maketitle

\section{Introduction}
Many thermodynamic and transport phenomena in superconductivity can
be  explained on the basis of a macroscopic order parameter, as
shown by the phenomenological approach of Ginzburg and Landau (GL),
proposed many year before Leon Cooper found that a weak attraction
binds electrons near to the Fermi surface to form pairs. From the GL
theory A. A. Abrikosov predicted the existence of two types of
superconductors according to the dimensionless coupling $\kappa$,
thus proving that this theory is well suited to describe alloys,
which are $\kappa > 1/\sqrt{2}$ superconductors, able to sustain a
stable vortex state in presence of an external applied
field~\cite{brandt}. Central to the present analysis is the
observation that at the core of Abrikosov's treatment lives a
condition which is totally independent of the Ginzburg-Landau free
energy. This condition is the requirement that the macroscopic order
parameter fulfils a ground state equation and from it the
supercurrent and the superfluid densities are shown to be directly
related to each other. It also follows from this condition that the
magnetization is proportional to the spatial average of the
superfluid density, a mean field result that remains valid in the
presence of thermal fluctuations~\cite{rosenstein}. Therefore the
ground state condition for the one-component order parameter (1COP)
superconductor lives at a more fundamental level than the
Ginzburg-Landau free energy expansion. To attain full predictive
power from the magnetization expression the superfluid density must
be determined in terms of fundamental parameters, and, at this
point, the free energy considerations becomes valuable, as shown by
Abrikosov, who obtained from the GL theory the magnetization as a
function of the temperature T and of the external applied magnetic
field, H, taken to be near to the upper critical field
$H_{c2}$~\cite{deGennesbook}.

In this letter we show that for the two-component order parameter
(2COP) superconductor the requirement that the macroscopic order
parameter fulfils a ground state equation leads to just two
possibilities, namely, the Abrikosov ground state, applied
individually to each one of the components, and a new ground state
that describes a textured superconducting layer because of the intrinsic phase
difference of $\pi$ between its distinct regions. Therefore we find here
that for the 2COP superconductor the ground state condition also
exists and lives in a more fundamental level than the free energy
expansion. In case the local field is equal to the applied one, the
Abrikosov ground state locks the 1COP into the lowest Landau level,
whereas this new 2COP ground state takes contributions from all
Landau levels.

In 1976 E. Bogomolny~\cite{bogomolny,rebbi,lukyanchuk01,lukyanchuk02} found,
while working in string theory, extra properties for the GL theory
valid for $\kappa=1/\sqrt{2}$: first order equations, instead of the
second order variational equations (Amp\`ere's law and the GL
equation) solve the GL theory. The Bogomolny's first order equations
share a common feature with the Abrikosov's treatment of the GL
theory: one of these equations is exactly the Abrikosov ground state
condition.

We also study here a 2COP free energy because the new ground state
condition is one of its Bogomolny equations. This means that at a
particular coupling, defined to be $\kappa=1/\sqrt{2}$, like for the
1COP case, the minimum of this 2COP free energy is reached by first
order equations instead of second order ones, which are then defined
as 2COP Bogomolny equations. Therefore this 2COP free energy theory
is a unique generalization of the 1COP GL theory. Previously
proposed 2COP GL theories~\cite{das88,sigrist91,fetter95,joynt02,babaev} do not
contain this feature, namely, none of them is minimized by the
present Bogolmony equations associated to this new ground state. We
stress that the study of this 2COP free energy is just complementary
to our major claim of a ground state condition independent of a free
energy proposal.

Nowadays a broad range of experimental data is pointing towards
common features for the superconductors whose pairing originates in
two-dimensional layers, such as the existence of two fundamental
energy scales~\cite{hufner,annette10} and the competition of the
superconducting state with a magnetic collective
state~\cite{sachdev09}. This leads to akin properties for the
layered superconductors albeit their distinct atomic structure. The
most studied materials are the cuprates whose two-dimensional layers
are made of copper and oxygen atoms~\cite{kohsaka,stephen,haug}.
Recently the pnictides brought excitement into the field because
superconductivity originates in two-dimensional layers made of iron
and arsenic atoms instead~\cite{chuang,drew,yildirim}.  The origin
of superconductivity, the microscopic nature of the pairing
mechanism~\cite{berg09,norman}, and the intrinsic magnetism are
under current
investigation~\cite{hufner,kohsaka,stephen,haug,chuang,drew,yildirim}.
We propose here that the new 2COP ground state will help to
understand the macroscopic properties of the cuprates and of the
pnictides because of its intrinsic layered structure.

The new ground state is described by a 2COP, $\Psi \equiv \left(
\vert \psi_1\vert \exp{(i\theta_1)},\; \vert \psi_2\vert
\exp{(i\theta_2)} \right)$, whose response to the presence of an
applied field $\vec H = H \hat x_3$ is a local magnetization
proportional to $\Psi^{\dagger}\vec \sigma \Psi$. In components this
magnetization is given by,
\begin{eqnarray}\label{m2}
M_1&=& -(\hbar q/mc)\langle \cos{\theta}|\psi_1||\psi_2|\rangle,\\
M_2&=& (\hbar q/mc)\langle \sin{\theta}|\psi_1||\psi_2|\rangle,\;\mbox{and}, \;\\
M_3&=& -(\hbar q/2mc)\langle |\psi_1|^2-|\psi_2|^2\rangle,
\end{eqnarray}
where $\theta=\theta_1-\theta_2$ is the phase difference between the
two order parameters and $\langle \cdots \rangle$ means spatial
average. Therefore there is a {\it local} transverse magnetization
to the applied field, but the total average magnetization should
vanish, $M_1=M_2=0$, to avoid a spontaneous torque in the system.
This is a novel feature, not present in the Abrikosov ground state,
whose magnetization is fully oriented along the applied field. The
key ingredient to make the {\it average} in plane magnetization to
vanish is to introduce texture, which means that the {\it local} transverse magnetization flips sign
by the phase shift $\theta \rightarrow \theta+\pi$.
Phase texture in two-gap superconductors has been discussed in the literature~\cite{gurevich06}.
It introduces interesting features such as the onset of solitons~\cite{tanaka01, gurevich03}
and an intercomponent Josephson interaction~\cite{doria88} that may directly relate to an abnormal AC loss peak~\cite{tanaka07}.

The 1COP GL theory has been applied to the cuprates either by including a mass
anisotropy or by the Lawrence-Doniach model, which treats
superconducting layers coupled by the Josephson effect~\cite{brandt}.
The present 2COP free energy allows for such
generalizations that will be considered elsewhere.

\section{One-component order parameter (1COP) ground state}
On the basis that the superconducting state can be described by a
macroscopic wavefunction, $\psi$, the supercurrent density is $\vec
J = (q/2m)\left( \psi^*\vec D \psi + c.c \right)$, $\vec D =
(\hbar/i)\vec \nabla - (q/c)\vec A$. Without invoking the GL theory,
Amp\`ere's law, $\vec \nabla \times \vec h = (4\pi/c)\vec J$, is
exactly solved for the local field, $\vec h=\vec \nabla \times \vec
A$, in terms of $\psi$ by assuming the ground state condition,
\begin{equation}
 D_{+}\psi=0, \label{con1}
\end{equation}
where $D_{\pm}=D_1\pm i D_2$. This ground state condition applies
for a bulk superconductor with continuous axial symmetry along the
applied field direction ($\vec H = H \hat x_3$), such that fields
only dependent on the coordinates orthogonal to it, $(x_1, x_2)$.
From the real and imaginary parts of Eq.(\ref{con1}), we find that
the supercurrent and superfluid densities are related by,
\begin{equation} \label{j1}
\vec J = -\frac{\hbar q}{2m}\vec \nabla \times \left(|\psi|^2\hat
x_3 \right)
\end{equation}
Integration of Amp\`ere's law gives that the local field is,
\begin{equation}\label{h31}
h_3=H-\frac{hq}{mc}|\psi|^2,\, \mbox{and} \quad 4\pi M =
-\frac{hq}{mc}\langle|\psi|^2\rangle.
\end{equation}
is the  magnetization. For the 1COP case the spatial average value
is over the orthogonal coordinates, such as for the magnetic
induction, $\vec B= \vec H+4\pi \vec M$, which becomes $\vec B
\equiv \langle h_3\rangle \hat x_3=(\int d^2x h_3/S)\hat x_3$, since
along the third direction and under translational invariance, it
suffices to consider integration over the unit cell area S. For a
constant field ($A_1=-Hx_2$, $A_2=0$), Eq.(\ref{con1}) is just the
lowest Landau level condition whose solution is
\begin{equation}\label{lll}
\psi = \sum_k c_k e^{ik x_1 -\frac{qH}{2\hbar c}\left(x_2+\frac{\hbar ck}{qH}\right)^2}.
\end{equation}

The set of wavenumbers $k$ and the constants $c_k$ are determined by
imposing periodic conditions to the order parameter and fixing the
number of vortices within the unit cell area. One obtains that $B =
N \Phi_0 /S$, for $N$ vortices within the cell. Notice that the
order parameter of Eq.(\ref{lll}) together with the local magnetic
field of Eq.(\ref{h31}) stem just from the ground state condition.
Nevertheless they describe the vortex state without invoking the
free energy, that only enters to select the vortex state of lowest
free energy. As shown by E.H. Brandt~\cite{brandt}, the order
parameter of Eq.(\ref{lll}) and the local field of Eq.(\ref{h31})
provide an excellent description of the full GL free energy solution
for fields in the range $0.5 H_{c2} \leq H \leq H_{c2}$. Therefore
the ground state condition of Eq.(\ref{con1}) does describe the GL
theory for an applied field not necessarily near to $H_{c2}$.

\section{Two-components order parameter (2COP) ground state}
Along the crystal's major axis, where the mass tensor is diagonal,
and an appropriate combination of the two order parameters is taken,
the supercurrent is expressed by $\vec J = \vec J_1+\vec J_2$, $J_j
= (q/2m)\left( \psi_j^*\vec D \psi_j + c.c \right)$, $j=1,\;2$. The
Abrikosov ground state assumes translational invariance along the
applied field such that the condition of Eq.(\ref{con1}) applies to
both components, and the local field becomes
$h_3=H-(hq/mc)(|\psi_1|^2 + |\psi_2|^2)$. This new ground state
satisfies the condition,
\begin{equation}\label{con2}
\vec \sigma \cdot \vec D \Psi=0, \; \mbox{where}\quad \Psi=
\left(\begin{array}{c} \psi_{1} \\ \psi_{2}  \\ \end{array} \right),
\end{equation}
where $\vec \sigma$ are the Pauli matrices, the generators of the
SU(2) group: $\sigma_1 {\scriptstyle =\left(
\begin{array}{cc} 0 & 1 \\ 1 & 0 \end{array} \right)}$, $\sigma_2
{\scriptstyle =\left(\begin{array}{cc} 0 & -i \\ i & 0 \end{array}
\right)}$, and  $\sigma_3 {\scriptstyle=\left( \begin{array}{cc} 1 &
0 \\ 0 & -1 \end{array} \right)}$. These matrices live on a vector
representation of the O(3) group, and so this ground state contains
an orientation and a direction in space, associated to the physical
superconducting layer linked to the crystal's major axes. Take the
two components of Eq.(\ref{con2}), $D_{3}\psi_1+D_{-}\psi_2=0$ and
$D_{+}\psi_1-D_{3}\psi_2=0$ to separate the two order parameters in
second order equations: $[(D_1^2+D_2^2)+D_3^2-(q\hbar/c)h_3]\psi_1 =
0 $ and $[(D_1^2+D_2^2)+D_3^2+(q\hbar/c)h_3]\psi_2 = 0 $. We show
that this new 2COP ground state is truly three-dimensional, although
it describes fields that evanesce from a layer, opposite to the
Abrikosov ground state, which is two-dimensional and describes a
translational invariant bulk state along the third direction. We
obtain the exponential decay away from the layer in presence of $H$
such that the local field is dominated by the applied field, $h_3
\approx H>0 $. The kinetic energy within the orthogonal plane is
positive, $(D_1^2+D_2^2)\psi_i \ge 0 $, $i=1,2$, and so, we must
have that $D_3^2\psi_2 \le 0$ in order to obtain a non-trivial
solution for the second equation. We take the gauge $A_3=0$ and that
the layer is at $x_3=0$, such that $\psi_i =
\exp{(-|q_3|x_3)}\psi_i(x_1,x_2)$, $i=1,2$, to obtain that,
\begin{eqnarray}
\left( D_{1}^2 + D_{2}^2 \right ) \psi_1 &=&  \big[ \left(\hbar q_3\right)^2+ \frac{\hbar c}{q}h_3 \big ] \psi_1 \label{eqq12}\\
\left( D_{1}^2 + D_{2}^2 \right ) \psi_2 &=&  \big[ \left(\hbar q_3\right)^2- \frac{\hbar c}{q} h_3 \big ] \psi_2,\label{eqq22}
\end{eqnarray}
For the case $h_3=H$ the system admits the general solution,
\begin{eqnarray}
\Psi &=& \sum_{n=0}^{\infty}C_n e^{-q_3(n)|x_3|}\left(
\begin{array}{c} \psi_{(n)}(x_1,x_2) \\ \psi_{(n-1)}(x_1,x_2)
\end{array} \right),  \label{expan1} \\ q_3(n)&=&\sqrt{2n\frac{q H}{\hbar c}},\; \psi_{(n-1)} = -\frac{i}{\hbar q_3(n)}D_{+}\psi_{(n)},\label{expan2}
\end{eqnarray}
where $(D_1^2+D_2^2)\psi_{(n)}=(\hbar q H/c)(2n+1)\psi_{(n)}$ and
$\psi_{(-1)}=0$ by definition. The first term retrieves the lowest
Landau level condition for the 1COP ($q_3=0$, $\psi_2=0$ and
$D_{+}\psi_1=0$) and so the above 2COP contains both two and three
dimensional contributions. We stress that each pair ($\psi_{(n)}$,
$\psi_{(n-1)}$) is bound by the same coefficients $c_k$ and wave
numbers $k$, but not necessarily two different pairs.

The major result of this letter is that under the ground state
condition of Eq.(\ref{con2}), the supercurrent becomes
\begin{equation} \label{j2}
\vec J = -\frac{\hbar q}{2m}\vec \nabla \times \left
(\Psi^{\dagger}\vec \sigma \Psi\right ).
\end{equation}
Integration of Amp\`ere's law gives that,
\begin{eqnarray}\label{hv2}
& \vec h = \vec H  - \frac{hq}{mc}\Psi^{\dagger}\vec \sigma \Psi,
\;\mbox{and }\quad 4\pi \vec M = - \frac{hq}{mc} \Big\langle
\Psi^{\dagger}\vec \sigma \Psi \Big\rangle,
\end{eqnarray}
for the local magnetic field and for the magnetization,
respectively. The latter is a three-dimensional
average value defined as $\vec B \equiv \langle \vec h \rangle= \int d^3x \vec
h/V$  for the magnetic induction, where V is the volume of the unit cell.

In conclusion several  ground states with elaborate magnetic
patterns can be constructed from our explicit 2COP solution $\Psi$,
given by Eqs.(\ref{expan1}) and (\ref{expan2}). As previously
discussed these ground states are textured which means that they have neighbor regions phase
separated by $\pi$ in order to render that $\Big\langle \Psi^{\dagger}
\sigma_1 \Psi \Big\rangle=$ $\Big\langle \Psi^{\dagger}\sigma_2 \Psi
\Big\rangle=0$, yielding a vanishing magnetization along the layer.
This in plane magnetization must vanish inside the bulk
of the superconductor, and there are several ways to do it, either
by considering two distinct layers with opposite in plane
magnetization, stripes~\cite{berg09} or checkerboard
patterns~\cite{valenzuela} inside a single layer.

\section{Virial relation}
We provide another derivation of Eqs.(\ref{h31}) and (\ref{hv2})
from the ground state conditions, Eqs.(\ref{con1}) and (\ref{con2}).
This derivation brings insight into the kinetic energy and is based
on the scalar virial relation~\cite{doria89,doria96,doria08} instead
of the supercurrent density. The scalar virial relation holds for
both the 1COP and the 2COP and states that the applied field times
the magnetic induction is proportional to the average kinetic energy
plus twice the average field energy:
\begin{eqnarray}\label{vt}
\frac{\vec H \cdot \vec B }{4\pi}= \Big\langle  \frac{|\vec D
\Psi|^2}{2m}\Big\rangle+2\Big\langle\frac{\vec
h^2}{8\pi}\Big\rangle,
\end{eqnarray}
The kinetic energy is expressed along the crystal's major axis,
taken appropriate combinations of the two components, that bring it
to the above form. We briefly review the 1COP case, considered by U.
Klein and B. P\"ottinger~\cite{klein} sometime ago. The virial
relation is $H B/ 4\pi= \langle|D_1 \psi|^2+|D_2 \psi|^2 \rangle
/2m+ \langle h_3^2\rangle/4\pi$, and from it, we obtain
Eq.(\ref{h31}). To show this, express the kinetic energy in terms of
the ground state condition of Eq.(\ref{con1}): $\sum_{j=1}^2
|D_j\psi|^2 = \sum_{j=1}^2 \nabla_j \left( i\psi^{*}D_j\psi \right)+
|D_{+}\psi|^2+(\hbar q/c)h_3|\psi|^2$. The total derivative term
does not contribute to the average kinetic energy because of
periodic boundary conditions. One obtains that $H
\langle{h_3}\rangle/ 4\pi= (\hbar q /2 m
c)\langle{h_3|\psi|^2}\rangle+ \langle{h_3^2}\rangle/4\pi$. Then one
easily verifies that the solution for this relation is indeed given
by Eq.(\ref{h31}). Similarly the average kinetic energy of
Eq.(\ref{vt}) can be expressed through the ground state condition,
using that,
\begin{eqnarray}\label{k2}
\Big\langle \vert \vec D \Psi \vert^2 \Big \rangle =\Big\langle
\vert \vec \sigma \cdot \vec D \Psi \vert^2\Big\rangle+ (\hbar
q/c)\Big\langle \vec h \cdot \Psi^{\dagger}\vec \sigma \Psi
\Big\rangle,
\end{eqnarray}
The virial relation of Eq.(\ref{vt}) leads to the local field of
Eq.(\ref{hv2}), by introducing Eq.(\ref{con2}) into the
above expression.

The kinetic energy density is intrinsically positive, and so,
helpful to understand why in presence of $H$ the 2COP has the top
component in a Landau level higher than the bottom one, according to
Eq.(\ref{expan1}). Take a constant applied field along the third
direction in Eq.(\ref{con2}), use Eqs.(\ref{hv2}) and (\ref{k2}),
and neglect the fourth order term under the assumption of weak order
parameter because of the proximity to the normal state. It follows
that $\langle \vert \vec D \Psi \vert^2\rangle/(\hbar q/c)\approx
H\langle \vert\psi_1\vert^2- \vert\psi_2\vert^2\rangle$, indicating
that $\langle \vert\psi_1\vert^2\rangle \ge
\langle\vert\psi_2\vert^2\rangle$. An interesting property is that
$\langle\Psi^{\dagger}\sigma_3\Psi\rangle=0$ holds for each pair
$\Psi$ given by $(\psi_{(n)},D_{+}\psi_{(n)}/i\hbar q_3(n) )$.

\section{Ginzburg-Landau Theory}
The 2COP GL theory with Bogomolny solution has three temperature
independent parameters, $\alpha_i(T)$, $i=1,2,3$, that form a vector
and consequently introduce intrinsically preferred directions for
the superconductor. Nevertheless it has a single dimensionless
coupling that for $\kappa=1/\sqrt{2}$ admits a Bogomolny solution,
as shown below.

We briefly review the derivation of the Bogomolny solution for the
1COP GL theory, whose free energy is $F(\vec B,T)= \langle |\vec D
\psi|^2/2m-\alpha(T) |\psi|^2 +\beta|\psi|^4/2 +\vec h^2/8\pi
\rangle$, where $\alpha(T)=(T_c-T)$ near to the transition and
$\beta>0$, rendering possible to have a non-trivial solution $\psi$
in case that $\alpha(T)>0$. Expressing quantities in dimensionless
units based on the coherence length, $\xi(T)^2=
\hbar^2/2m|\alpha(T)|$, and the thermodynamic magnetic field,
$H_c(T)^2/4\pi= \alpha(T)^2/\beta$, one finds that the GL theory
only depends on a single dimensionless temperature independent
coupling $\kappa= H_{c2}(T)/(\sqrt{2}H_c(T))=(m c/\hbar
q)\sqrt{\beta/(2\pi)}$, the upper critical field being
$H_{c2}(T)=\Phi_0/2\pi\xi(T)^2$. Hereafter we use the dimensionless
quantities: ${\vec x} \rightarrow \vec x/\xi$ (${\vec
\nabla}\rightarrow \xi \vec \nabla$), $\vec A\rightarrow
(2\pi\xi/\Phi_0) \vec A$ ($\vec D \rightarrow (\hbar/\xi) \vec D$,
$\vec D \rightarrow \vec \nabla/i-\vec A$, $\vec h \rightarrow \vec
h / H_{c2}$ ($\vec H\rightarrow \vec H / H_{c2}$), $F\rightarrow
F/(H_c^2/4\pi)$ ($G\rightarrow G/(H_c^2/4\pi)$). For instance, the
thermodynamic relation $\vec H = 4\pi \partial F/
\partial \vec B $ becomes in reduced units
$\vec H = (1/2\kappa^2) \partial F / \partial \vec B $. Consider the
special case of symmetry along the third direction defined by $H$
and decompose the kinetic energy using the key condition to obtain
that the free energy is $F = \langle
|D_{+}\psi|^2+h_3|\psi|^2-|\psi|^2 +|\psi|^4/2 +\kappa^2 h_3^2
\rangle$. A straightforward rearrangement of the terms leads to
$F=-1/2+B+F_0$, where $ F_0 = \langle |D_{+}\psi|^2 \rangle+ \langle
\left ( h_3+ |\psi|^2-1 \right )^2 \rangle/2  + (\kappa^2 -
1/2)\langle h_3^2 \rangle$. For $\kappa=1/\sqrt{2}$ $F_0$ is
positive and hold the first order differential equations,
$D_{+}\psi=0$ and $h_3+ |\psi|^2-1 =0$, at the minimum since
$F_0=0$. Consequently, at this particular coupling, vortices do not
interact as the free energy is additive in their number, $N$, and
$F=-1/2+B$, where the magnetic induction is $B = N \Phi_0 / S$.

The 2COP GL theory which has a Bogomolny symmetry is,
\begin{eqnarray}\label{freea2}
F = \Big\langle \frac{\vert \vec D \Psi \vert^2}{2m}-\vec \alpha(T)
\cdot  \Psi^{\dagger}\vec \sigma \Psi + \frac{\beta}{2} \left (
\Psi^{\dagger}\Psi\right )^2+\frac {\vec h^2}{8\pi}\Big\rangle.
\end{eqnarray}
The out of plane coupling, $\alpha_3$, is associated to $T_c$,
whereas the in plane ones, $\alpha_1$ and $\alpha_2$, are related to
$\theta$ transitions. The simplest of all situations,
$\alpha_1=\alpha_2=0$, brings some insight into the problem since
the remaining quadratic coupling becomes $-\alpha_3(T) \left (
|\psi_1|^2 - |\psi_2|^2\right )$. In this case $\psi_1$ admits a
non-trivial solution below $T_c$ ($\alpha_3(T)>0$), whereas $\psi_2$
does not, and must vanish. However above $T_c$ ($\alpha_3(T)<0$) a
non-trivial solution for $\psi_2$ must be prevented and this is done
by assuming that $\alpha_3$ vanishes above $T_c$. We get further
insight into the properties of this theory by doing a SU(2) rotation
$\Psi_{R}=U(T)\Psi$, such that $\vec \alpha (T)\cdot U(T)\vec \sigma
U(T)^{\dagger}=|\vec \alpha(T)|\sigma_3$. Under this transformation
the free energy becomes ${\textstyle F= \langle \vert \vec D \Psi_R
\vert^2/2m- \vert \vec \alpha(T) \vert\Psi_R^{\dagger} \sigma_3
\Psi_R +  \beta( \Psi_R^{\dagger}\Psi_R )^2/2+}$ $+{\textstyle \vec
h^2/8\pi \rangle}$. According to the previous argument we conclude
that no stable second component is possible, namely,
${\psi_{R}}_2=0$ turning this theory essentially into a 1COP GL
theory for ${\psi_{R}}_1$. Notice however an important difference,
${\psi_{R}}_1$ does not vanish at $T_c$ because of the presence of
$\alpha_1$ and $\alpha_2$, assuming that these parameter only vanish
above $T_c$. The previously defined quantities $H_c(T)$, $\xi(T)$,
and $H_{c2}(T)$ depend on $|\vec \alpha(T)|$ instead and this leads
to the same dimensionless $\kappa$ of the 1COP case. For instance
the one-dimensional surface energy barrier for the 2COP theory of
Eq.(\ref{freea2}) is found to be $\vec \alpha$ independent and only
$\kappa$ dependent, exactly like for the 1COP case, because the
above rotation is possible. However for the three-dimensional
problem we demand from the theory that the ground state condition of
Eq.(\ref{con2}) applies to $\Psi$ and not to $\Psi_R$. This will
halt the rotation preventing the theory to become a one component
free energy, but keeping its single scale properties. Reduced units
are defined as before, and the free energy becomes,
\begin{eqnarray}\label{freeb2}
{\textstyle F = \Big\langle \vert \vec D \Psi \vert^2-\frac{\vec
\alpha}{|\vec \alpha|} \cdot  \Psi^{\dagger}\vec \sigma \Psi +
\frac{1}{2} \left ( \Psi^{\dagger}\Psi\right )^2+\kappa^2\vec
h^2\Big\rangle},
\end{eqnarray}
For zero magnetic field and constant order parameter the minimum of
the 2COP GL potential gives the values for the two physically
relevant order parameters. Using that $\left( \Psi^{\dagger} \Psi
\right)^2= \left( \Psi^{\dagger}\vec \sigma \Psi \right)^2 $, the GL
potential becomes, $F_P=-1/2+ \left( \Psi\vec \sigma \Psi - \vec
\alpha/|\vec \alpha|\right)^2/2$, whose minimum is reached at $
\vert \psi_1\vert^2 = (|\vec \alpha |+\alpha_3)/2|\vec \alpha|)$,
$\vert \psi_2 \vert^2 = (|\vec \alpha|- \alpha_3)/2|\vec \alpha |)$
and $\tan{\theta}=-\alpha_2/\alpha_1$. Notice that it is always true
that $|\psi_1| \ge |\psi_2| $, and that above $T_c$ they become
equal, assuming that $\alpha_1$ and $\alpha_2$ exist above the
transition. This restores a U(1) invariance to the theory. A change
of sign in $\alpha_1(T)$ or $\alpha_2(T)$ changes the phase $\theta$
but not the minimum of $|\psi_1|$ and $|\psi_2|$. The transformation
$\alpha_1\rightarrow -\alpha_1$ and $\alpha_2\rightarrow -\alpha_2$
is a symmetry of the minimum, equivalent to a $\theta \rightarrow
\theta+\pi$ rotation. This theory has a Bogomolny solution for
$\kappa=1/\sqrt{2}$, given by $\vec \sigma \cdot \vec D \Psi=0$ and
$ \vec h + \Psi^{\dagger}\vec \sigma \Psi -\vec \alpha/ |\vec
\alpha|=0$, as the free energy can be expressed by,
\begin{eqnarray}\label{freeb3}
F &=& \Big\langle \vert \vec \sigma \cdot \vec D \Psi \vert^2 + \big
[\vec h + \left( \Psi^{\dagger}\vec \sigma \Psi -\vec
\alpha(T)/|\vec \alpha(T)|\right) \big ]^2/2 \Big\rangle+
\nonumber \\
&+&\left(\kappa^2-1/2\right)\Big\langle\vec h^2 \Big\rangle +(\vec
\alpha(T) /|\vec \alpha(T)|)\cdot \vec B/2-1/2.
\end{eqnarray}
Thus the 2COP free energy studied here is distinct from all other
free energies studied so
far~\cite{das88,sigrist91,fetter95,joynt02,babaev} because the
ground state condition holds in two distinct applied field H and
$\kappa$ regimes, showing that it is a unique generalization of the
1COP GL theory: $\kappa
> 1/\sqrt{2}$ and $0.5 H_{c2} \leq H \leq H_{c2}$ (Abrikosov), and
also $\kappa=1/\sqrt{2}$ throughout the whole $H$ regime
(Bogomolny).

In conclusion, we have found here a textured ground state
for the two-component order parameter whose intrinsic transverse
magnetic moment averages to zero along the superconducting layer
because of $\pi$ phase difference between distinct regions. This
ground state solves one of the Bogomolny equations of a free energy
obtained here. The ground state condition lives in a more
fundamental level than that of the free energy expansion.

\acknowledgments We thank Ernst Helmut Brand and Said Salem
Sugui Jr. for helpful discussions. Part of this work was supported
by the bilateral project between Flanders and Brazil. M.M. Doria
also thanks CNPq, Capes and FAPERJ (Brazil).

\end{document}